\documentclass[aps,twocolumn,preprintnumbers,nofootinbib,superscriptaddress]{revtex4-2}
\usepackage{amsmath,amssymb,amsfonts,amsthm}
\usepackage{graphicx}
\usepackage{bm}
\usepackage{times}
\usepackage{xcolor}
\usepackage{orcidlink}
\usepackage{hyperref}

\hypersetup{colorlinks=true,
breaklinks=true,
pdfstartview=Fit,
linkcolor=magenta,
citecolor=blue,
urlcolor=blue}
\usepackage{float}

\newcommand{\be}{\begin{equation}}
\newcommand{\ee}{\end{equation}}
\newcommand{\ba}{\begin{eqnarray}}
\newcommand{\ea}{\end{eqnarray}}
\newcommand{\bal}{\begin{align}}
\newcommand{\eal}{\end{align}}

\newcommand{\dd}{\text{d}}

\newcommand{\bw}{\begin{widetext}}
\newcommand{\ew}{\end{widetext}}

\definecolor{darkgreen}{rgb}{0.3,0.7,0.1}

\begin{document}

\title{Comment on {\textquotedblleft Exact, nonsingular black holes from a phantom DBI field as primordial dark matter \textquotedblright}}

\author{Mustapha Azreg-A\"{i}nou\,\orcidlink{0000-0002-3244-7195}}
\email{azreg@baskent.edu.tr}
\affiliation{Ba\c{s}kent University, Engineering Faculty, Ba\u{g}l{\i}ca Campus, Ankara 06790, T\"{u}rkiye.}

\author{Marco A. A. de Paula\,\orcidlink{0000-0002-9808-7495}}
\email{marcodepaula@ufpa.br}
\affiliation{Faculdade de Ciências Naturais, Universidade Federal do Par\'a, Campus Universit\'ario do Tocantins-Cametá, 68400-000, Camet\'a, Par\'a, Brazil.}
\affiliation{Departamento de F\'isica, Universidade Federal da Para\'iba, 58051-970, Jo\~ao Pessoa, Para\'iba, Brazil.}

\begin{abstract}

We demonstrate that the metric and scalar configuration proposed in Ref.~\cite{dbi26} do not satisfy the complete set of field equations of the Dirac-Born-Infeld (DBI) field theory minimally coupled to general relativity on a radial domain containing the event horizon. The physical DBI branch for this gravitational setup terminates outside the event horizon. Thus, the proposed configuration does not describe a black hole solution of that theory.

\end{abstract}

\maketitle

Recently, in the framework of general relativity minimally coupled to a Dirac-Born-Infeld (DBI) scalar field, the authors of Ref.~\cite{dbi26} obtained an exact regular solution and claimed that it was a black hole (BH). We show that this solution is indeed regular, but it does not describe a BH, since the field equations are satisfied in a radial domain that does not include an event horizon. Throughout this comment, we use the notation of Ref.~\cite{dbi26}, except that we replace $\kappa^2$ with $\kappa$.

The action of general relativity minimally coupled to the DBI scalar field $\phi$ is given by
\begin{equation}
\label{eqLag} S=\int \dd ^4x\sqrt{-g}\Big(\dfrac{R}{2\kappa}+\epsilon\mathcal{L}_\phi \Big)\,,
\end{equation}
where $g$ is the determinant of the metric tensor $\textbf{g}$, $R$ is the Ricci scalar, $\kappa=8\pi G/c^4$, and $\epsilon = -1$ ($\epsilon = +1$) for a normal (phantom) scalar field. Moreover, $\mathcal{L}_\phi$ is the Lagrangian density associated with the scalar field, namely,
\begin{equation}\label{eqL}
\mathcal{L}_\phi=V(\phi)\Big(1-\sqrt{1+\dfrac{2X}{\Lambda^4}}\,\Big)\,,\; X\equiv -\frac{1}{2}g^{\mu\nu}\nabla_{\mu}\phi\nabla_{\nu}\phi\,.
\end{equation}
in which $\Lambda$ is the intrinsic energy scale, $V(\phi)$ is the DBI potential function. The equations of motion assume the form~\cite{dbi26}
\begin{align}
&\label{eqphi}
\nabla_{\mu}\Big(\frac{V(\phi)\nabla^{\mu}\phi}{Q}\Big) =  
\Lambda^{4}V'(\phi)(1-Q)\,,\\
\label{eqgr}
&G^{\mu}_{\nu} = \kappa\epsilon
\Big[\mathcal{L}_\phi \delta^{\mu}_{\nu} 
- {\frac{V(\phi)(\nabla^{\mu}\phi)(\nabla_{\nu}\phi)}{\Lambda^4 Q}}\Big]=\kappa T^{\mu}_{\nu}\,,
\end{align}
with
\begin{equation}
\label{Qofr}Q\equiv \sqrt{1+\frac{2X}{\Lambda^4}}> 0\,.
\end{equation}
In this work, the prime notation ($^{\prime}$) implies a derivative with respect to the argument shown between parentheses. Consequently, $\phi '(r)=\dd\phi/\dd r$ and $\phi '(t)=\dd\phi/\dd t$.

Choosing the metric ansatz of the form
\begin{equation}\label{eqm1}
\dd s^2=-c^2 f(r)\dd t^2+\frac{\dd r^2}{f(r)}+\rho^2(r)\dd\Omega^2\,,
\end{equation}
Eqs.~\eqref{eqphi} and $G^{t}_{t}=G^{\theta}_{\theta}$ reduce to
\begin{align}
&\label{eqphi2}
\frac{1}{\rho^2(r)} \frac{\dd}{\dd r}\Big[\frac{\rho^{2}(r)f(r)\phi'(r)V(\phi)}{Q}\Big] = \Lambda^4V'(\phi)(1-Q) \,,\\
\label{eqgrttpp}
&\rho^2(r) f''(r)-2 \Big[\big(\rho '(r)\big)^2+\rho (r) \rho ''(r)\Big] f(r)+2=0\,,
\end{align}
respectively. In Ref.~\cite{dbi26}, $\rho(r)$ was chosen $\rho(r) = \sqrt{r^2+a^2}$, and it yields the function $f(r)$ upon solving~\eqref{eqgrttpp}. Thus, we get
\begin{equation}\label{eqf}
f(x)=1+3x_g\Big[x-(1+x^2)\arctan\Big(\frac{1}{x}\Big)\Big]\,,
\end{equation}
where $a$ is a regularization parameter, and we define
\begin{equation}\label{eqxg}
x\equiv \frac{r}{a}\quad\text{and}\quad	x_g\equiv \frac{GM}{ac^2}\,.
\end{equation}

Now, from Eq.~\eqref{eqgr}, we obtain the following equations:
\begin{align}
\label{eqgtt}&G^{t}_{t}=\epsilon\kappa V(\phi)[1-Q(r)]\,, \\
\label{eqgttgrr}&G^{t}_{t}-G^{r}_{r}=\epsilon\kappa V(\phi)[1-Q(r)]\dfrac{[1+Q(r)]}{Q(r)}\,.
\end{align}
In terms of $f(r)$ and $\rho(r)$, we obtain
\begin{equation}\label{eqgttgrr3}
G^{t}_{t}-G^{r}_{r}= 2~\frac{\rho ''(r)}{\rho (r)}~f(r)\,,
\end{equation}
with $\rho''(r)=a^2/\rho^3>0$. We can cancel $f$ from both sides of Eq.~\eqref{eqgttgrr}, assuming that $f \neq 0$ in order to show that
\begin{equation}\label{eqV0}
2~\frac{\rho ''(r)}{\rho (r)}=\epsilon\kappa V(\phi)~\frac{\big(\phi'(r)\big)^2}{\Lambda^4 Q(r)}\,.
\end{equation}
At the points where $\big(\phi '(r)\big)^2$ remains positive and finite, all we can conclude from Eq.~\eqref{eqV0} is that $ \epsilon V(\phi)/Q$ is finite, i.e.,
\begin{equation}\label{eqV}
0 < \epsilon \frac{V(\phi)}{Q(r)} <\infty\,.
\end{equation}
Since $Q>0$, Eq.~\eqref{eqV} implies $\epsilon V>0$. Yet, we emphasize a point concerning the sign of the matter sector. In
Eq.~\eqref{eqLag}, $\epsilon$ and $V$ enter only through their product. The low-energy kinetic term is $\epsilon \mathcal{L}_\phi=-(\epsilon V(\phi)/\Lambda^4)X+O(X^2)$. Therefore, the combinations $(\epsilon,V)=(-1,V<0)$ and $(\epsilon,V)=(+1,V>0)$ describe the same phantom-sign kinetic term;
the former is not a standard scalar field. Moreover, Eq.~\eqref{eqV} constrains only the ratio $\epsilon V(\phi)/Q(r)$. Consequently, if $V$ and $Q$ vanish simultaneously, the ratio may have a finite positive limit even though the required condition $Q>0$ fails at that point.

The ratio of Eq.~\eqref{eqgttgrr} to Eq.~\eqref{eqgtt} yields
\begin{equation}\label{eqsQ}
Q(r)=-\frac{G^{t}_{t}}{G^{r}_{r}}\,.
\end{equation}
As we shall see later, the ratio $G^{t}_{t}/G^{r}_{r}$ \emph{does not have a constant sign}, and since we want $Q\geq 0$, this implies that the solution \emph{may not be valid} on the whole half real axis $[0,\,\infty)$. From Eqs.~\eqref{Qofr},~\eqref{eqgtt}, and~\eqref{eqsQ}, we find that
\begin{align}
\label{eqsV}&\epsilon V(\phi)=\frac{G^{t}_{t}G^{r}_{r}}{\kappa(G^{t}_{t}+G^{r}_{r})}\,,\\
\label{eqsF}&\frac{\big(\phi'(r)\big)^2}{\Lambda^4}=\frac{1-Q^2}{f}=\frac{1}{f}\Big[1-\Big(\frac{G^{t}_{t}}{G^{r}_{r}}\Big)^2\Big]\,.
\end{align}
Here again, the sign of $G^{t}_{t}G^{r}_{r}/(G^{t}_{t}+G^{r}_{r})$ is not constant.

It was stated in Ref.~\cite{dbi26} that for
\begin{equation}\label{eqBH}
M>M_{\text{relic}}\equiv\frac{2 a c^{2}}{3\pi G}\Longrightarrow x_g>x_{\text{BH}}\equiv\frac{2}{3\pi }\,,
\end{equation}
the solution~\eqref{eqf} is a non-extremal BH spacetime. We show that this statement is not correct. From Eq.~\eqref{Qofr}, we see that if $\big(\phi '(r)\big)^2$ remains finite, then the functions $f$ and $1-Q$ must have the same sign if $f(r)$ corresponds to a BH, \emph{provided that} $Q$ \emph{remains real and positive}. In this case, when $f$ changes sign at the horizon $r_h$, $1-Q$ also changes sign, and the limit of their ratio remains finite (this applies to $f$ and $1-Q^2$ as well). In fact, this was justified graphically in the upper plot of Fig. 2 in Sec. IV of the Supplemental Material of Ref.~\cite{dbi26}, where $(1-Q^2)/f$ is shown to maintain a positive sign. Yet, at the horizon, the combination of Eqs.~\eqref{eqgttgrr3} and~\eqref{eqsQ} leads to
\begin{equation}
\label{product}G^{t}_{t}(r_{h})[1+Q(r_{h})] = 0.
\end{equation}
Thus, we have two options, namely, either $G^{t}_{t}(r_{h}) = 0$ or $Q(r_{h}) = -1$. The function $G^{t}_{t}(r_{h})$ can be written as 
\begin{equation}
G^{t}_{t}(r_{h}) = \dfrac{r_{h} f^{\prime}(r_{h})-1}{a^{2}+r^{2}_{h}}.
\end{equation}
Solving $f(r_{h}) = 0$ for $x_{g}$ and inserting the resulting expression into the equation above, one can show that it implies that for every horizon radius $r_{h} > 0$, we have $G^{t}_{t}(r_{h}) < 0 $. Therefore, the product can vanish only through the second option, yielding $Q(r_{h}) = -1$. However, Eq.~\eqref{Qofr} implies that $Q(r) > 0$ holds throughout the claimed radial domain. Consequently, Eq. \eqref{product} can be satisfied only if $G^t_t(r_h) = 0$. This contradicts $G^t_t(r_h) < 0$. Hence, the field equations cannot be satisfied at $r_h$, and the solution cannot describe a BH.

We illustrate the above conclusions in Fig.~1 of the Supplemental Material. Using Eqs.~\eqref{eqsQ}--\eqref{eqsF} and Eq.~\eqref{eqf}, we display $2a^2(1+x^2)^2G^{t}_{t}$, $\big(\phi'(r)\big)^2/(2\Lambda^4)$, $f(r)$, $Q(r)$, and $8a^2\pi\kappa\epsilon V$ for $x_g=4/(3\pi)=2x_{\text{BH}}$ ($M=2M_{\text{relic}}$) [cf. Eq.~\eqref{eqBH}].

For sufficiently large values of $x_g$, as illustrated in Fig.~2 of the Supplemental Material, $\big(\phi'(r)\big)^2$ diverges at finite radii $r_s$ where $G^r{}_r(r_s)=0$. Near such a zero, we have $\big(\phi'(r)\big)^2\propto(r-r_s)^{-2}$, so the scalar field itself diverges logarithmically at finite $r$. This first occurs at $x_g\simeq0.989\simeq4.66x_{\text{BH}}$. However, these divergences occur inside the horizon, where the scalar gradient is timelike ($X>0$). Therefore, they do not violate the DBI square-root reality bound.

In summary, the metric function~\eqref{eqf} solves the reduced geometric equation, but the complete DBI field equations are not satisfied on a radial domain containing the horizon. The obstruction is the sign condition $Q>0$: the horizon requires $Q=-1$, and the physical exterior branch ends at $r_*>r_h$ with $Q(r_*)=0$. The result concerns the specific action~\eqref{eqLag}. Moreover, since $Q\to 1$ asymptotically and $Q(r_h)=-1$, the function $Q(r)$ vanishes at $r_*$ where $f(r_*) > 0$, meaning that the valid exterior domain where $Q > 0$ terminates before reaching the horizon $f(r_h) = 0$.

As a final comment, for the generalized ansatz $\rho(r)=(r^p+a^p)^{1/p}$ considered in Ref.~\cite{dbi26}, asymptotically flat solutions of Eq.~\eqref{eqgrttpp} can be constructed for any $p>1$, provided that $f(r)$ is nonconstant. However, boundedness of $\rho''(r)$ at $r=0$ requires $p\geq2$, while smooth radial continuation through $r=0$ requires even $p$. These qualifications distinguish asymptotic flatness from regularity at the center. Further details on these issues are outside the scope of this Comment and will be investigated elsewhere~\cite{other}.

M. A. A. de Paula acknowledges the Conselho Nacional de Desenvolvimento Cient\'ifico e Tecnol\'ogico (CNPq) from Brazil for partial financial support under the Grant No. CNPq/PDJ 150589/2025-5.

\section*{Data availability}

There are no publicly available research data or software supporting this manuscript. Requests for further information or data should be sent to the authors.

\clearpage
\section*{Supplemental Material}

In this Supplemental Material, all equation numbers and citations refer to the Comment. 

\begin{figure}[H]
	\centering
	\includegraphics[width=\columnwidth]{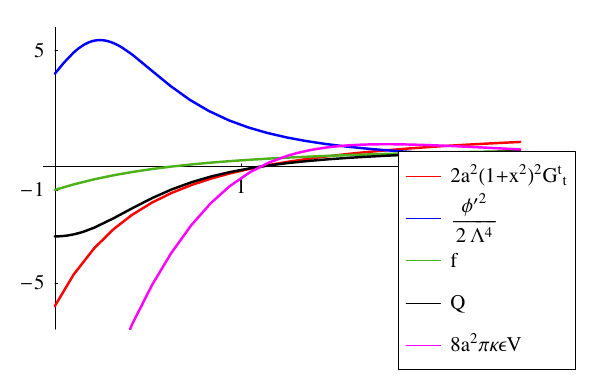}
	\caption{\footnotesize{Relevant functions $2a^2(1+x^2)^2G^{t}_{t}$, $\big(\phi'(r)\big)^2/(2\Lambda^4)$, $f(r)$, $Q(r)$, and $8a^2\pi\kappa\epsilon V$}, as a function of $x=r/a$, corresponding to the solution given by Eq.~\eqref{eqf}. We took $x_g=4/(3\pi)=2x_{\text{BH}}$ ($M=2M_{\text{relic}}$). We have $G^{t}_{t}(1.10278)=Q(1.10278)=V(1.10278)=0$ and $f(0.621184)=0$.}\label{Fig1c}
\end{figure}

In Fig.~\ref{Fig1c}, using Eqs.~\eqref{eqsQ}--\eqref{eqsF} and Eq.~\eqref{eqf} of the Comment, we display $2a^2(1+x^2)^2G^{t}_{t}$, $\big(\phi'(r)\big)^2/(2\Lambda^4)$, $f(r)$, $Q(r)$, and $8a^2\pi\kappa\epsilon V$ for $x_g=4/(3\pi)=2x_{\text{BH}}$ ($M=2M_{\text{relic}}$) [cf. Eq.~~\eqref{eqBH}]. We see that $Q<0$ (and $\epsilon V<0$) for $0\leq x<1.10278$, which implies that the solution is not valid for $0\leq x<1.10278$ and shows that $G^{t}_{t}/G^{r}_{r}$ does not have a constant sign. We also observe that at the point where $f$ vanishes, $x=0.621184$, both $\big(\phi'(r)\big)^2$ and $\epsilon V$ are finite, but $G^{t}_{t}(x=0.621184)\neq 0$, showing that Eq.~\eqref{eqf} does not satisfy Eq.~~\eqref{eqgtt}. This is because at $x=0.621184$, we have $Q(0.621184)=-1$ instead of $Q(0.621184)=+1$. The authors of Ref.~~\cite{dbi26} have checked in Fig. 2 of their Supplemental Material the sign of $1-Q^2$ against that of $f$, and should have checked the sign of $1-Q$ against that of $f$ instead.

\begin{figure}[H]
	\centering
	\includegraphics[width=\columnwidth]{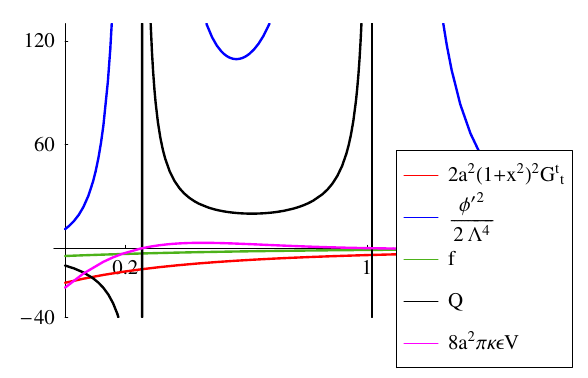} \includegraphics[width=\columnwidth]{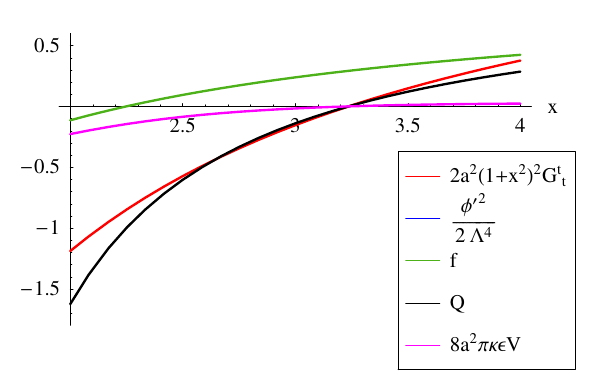}
	\caption{\footnotesize{Top panel: Relevant functions $2a^2(1+x^2)^2G^{t}_{t}$, $\big(\phi'(r)\big)^2/(2\Lambda^4)$, $f(r)$, $Q(r)$, and $8a^2\pi\kappa\epsilon V$}, as a function of $x=r/a$, corresponding to the solution Eq.~\eqref{eqf}. Bottom panel: Same as the top panel, but for $x \in [2,4]$. We took $x_g=11/(3\pi)=5.5x_{\text{BH}}$ ($M=5.5M_{\text{relic}}$). Moreover, we have $G^{t}_{t}(3.23675)=Q(3.23675)=V(3.23675)=0$ and $f(2.24903)=0$.}\label{Fig2c}
\end{figure}

We have obtained similar graphs for other values of $x_g>x_{\text{BH}}$ where $Q$ and $\big(\phi'(r)\big)^2$ remain finite for all $x\geq0$. For higher values of $x_g$, specifically beginning at $x_g\simeq0.989\simeq4.66x_{\text{BH}}$, the scalar derivative develops divergences at finite radii $r_s$ where $G^r{}_r(r_s)=0$. For $x_g=11/(3\pi)=5.5x_{\text{BH}}$ ($M=5.5M_{\text{relic}}$), as shown in Fig.~\ref{Fig2c}, the quantities remain finite near the zero of $f(x)$ and diverge toward the center. At $x=2.24903$, where $f=0$, $G^t{}_t\neq0$ and $Q=-1$. The figure also shows the behavior in the outer-radius interval $x\in[2,4]$.

\end{document}